# $SU(3)_C \otimes SU(4)_L \otimes U(1)_X$ model with two Higgs 4-plet scalars

Sutapa Sen and Aparna Dixit

Department of Physics, Christ Church College, Kanpur 208001, India

## Abstract

In this work we present an extension of the Standard model as a local gauge group $SU(3)_C \otimes SU(4)_L \otimes U(1)_X$ embedded in $SU(3)_C \otimes SU(4)_L \otimes U(1)_{Y_0} \otimes U(1)_{B-L}$ gauge group. The $U(1)_{Y_0}$ gauge symmetry is introduced to obtain an anomaly-free three- generation model without exotic electric charges. A minimal scalar sector is considered with two Higgs 4-Plet representations of $SU(4)_L$ and a singlet scalar The 15-plet gauge bosons of $SU(4)_L$ contain two new neutral bosons which mix with Standard model W. The (B-L) charge is assigned to $U(1)_{B-L}$ gauge symmetry so that $SU(4)_L$ gauge bosons do not contain bileptons. The model is analyzed for mass spectrum of Higgses, fermions and gauge bosons with results similar to an economical 3-3-1 model with two Higgs triplets. In 3-4-1 model both 4-plet scalars contain two neutral components each .There is no lepton number violation in 3-4-1 case.

## I. Introduction

The Standard Model (SM) with gauge symmetry $SU(3)_C \otimes SU(2)_L \otimes U(1)_Y$ has been very successful in describing strong and electroweak interactions[1] .In spite of these successes, it fails to explain several fundamental issues like generation number problem, neutrino masses and oscillations, the origin of charge quantization, CP-violation etc. Some models address the problem of origin of generations from larger symmetries[2] as well from cancellations of quiral anomalies [3]

One of the simplest solution to these problems is to enhance the SM gauge symmetry to SU(3) and SU(4) groups. Over the last decade, extensions of the SM in which the gauge symmetry is $SU(3)_C \otimes SU(3)_L \otimes U(1)_X$ known as 331 models have been studied in literature[4-8]. These models offer attractive solution to the generation number problem which arise from cancellations of chiral anomalies and also predict charge quantization[9] and neutrino oscillations[10] There are two popular versions for the 3-3-1 models , of which one includes quarks with exotic charges  $-\frac{4}{3}$ and $\frac{5}{3}$ and double charged gauge bosons [4,5]. A second type includes additional quarks and gauge bosons without exotic charges[6,7].

An extension of SM gauge group to $SU(4)_L \otimes U(1)_X$ has been proposed [11,12,14,15] for three-family models without exotic charges .The Little Higgs models also deal with extended electroweak symmetry 3-3-1 and 3-4-1 around TeV scale [13]. The 3-3-1 model with exotic charges embedded in $SU(4)_C \otimes SU(4)_{L+R}$ has been studied recently [16].

A general expression for electric charge operator in the 3-4-1 model is [14]

$$Q = aT_{3L} + \frac{b}{\sqrt{3}}T_{8L} + \frac{c}{\sqrt{6}}T_{15L} + XI_4 \qquad (1)$$

where $T_i, i = 3,8,15$ are given by $T_{iL} = \frac{\lambda_{iL}}{2}$. The Gell-Mann matrices for $SU(4)_L, \lambda_{iL}$ are normalized as $\text{Tr}(\lambda_i \lambda_j) = 2\delta_{ij}$ and $I_4 = Dg(1,1,1,1)$ is the 4 x 4 unit matrix

The 3-4-1 models with weak isospin $SU(2)_L$ of SM embedded as $SU(2)_L \subset SU(4)_L$, and fermions in fundamental representations $4_L, \overline{4}_L$ have the parameter a = 1 [17]. The parameters b,c differ for the models with and without exotic charges.

A systematic study [17] of the $SU(3)_C \otimes SU(4)_L \otimes U(1)_X$ local gauge symmetry has been recently performed for both models with and without exotic charged particles. For three-family models with gauge bosons with electric charges $0, \pm 1$ only, the possibilities for simultaneous values of b,c are

b = c = 1; b = c = - 1; b = 1,c = -2 ; b = -1,c = 2.

The phenomenology for 3-4-1 models has been studied only for few cases. The case of a = 1, b = 1,c = -1 has been studied for 3-families in a 3-4-1 model without exotic charges [15].The case of b = 1, c = - 2 has been considered [14]and correspond to model **E** of Ref.[17] In this work we present a two-Higgs 4-plet 3-4-1 model with a = 1,b = -1 and c = 2 corresponding to model **F** [17]. This case for model without exotic electric charges has not been studied earlier. The covering group is $SU(3)_C \otimes SU(4)_L \otimes U(1)_{B-L} \otimes U(1)_{Y_0}$. It is interesting to note that $SU(8) \supset SU(3) \otimes SU(4) \otimes U(1)^2$ gauge group with extra $U(1)$ gauge symmetry has been considered in 7D orbifold GUT models [19].

We consider a 3-4-1 model as derived from 3-4-1-1 gauge symmetry with

$$U(1)_{B-L} \otimes U(1)_{Y_0} \longrightarrow U(1)_X \; ; \; X = \frac{(B-L)}{2} + Y_0 = \frac{(B-L)'}{2} \tag{2}$$

The new flavor $Y_0 = \pm \frac{1}{2}$ for 4-plet Higgses and singlet fermions.

.For a = 1, b = -1, c = 2 values

$$Q = T_{3L} - \frac{1}{\sqrt{3}} T_{8L} + \frac{2}{\sqrt{6}} T_{15L} + XI_4 \tag{3}$$

The diagonal generators of $SU(4)_L$ include

$$T_{3L} = \frac{1}{2} Dg(1,-1,0,0); T_{8L} = \frac{1}{2\sqrt{3}} Dg(1,1,-2,0); T_{15L} = \frac{1}{2\sqrt{6}} Dg(1,1,1,-3)$$

This is analogous to the electric charge operator in a 3-3-1 model [18,20] embedded in the larger 3-4-1 group with $X_N$ hypercharge given by $X_N = \frac{2}{\sqrt{6}} T_{15L} + XI_4$

$$Q = T_{3L} - \frac{1}{\sqrt{3}} T_{8L} + X_N \tag{4}$$

The case of two-Higgs triplet model (economical 3-3-1 model) comes under this category [18,20]. It is interesting to study a two Higgs four-plet model with ordinary electric charges which promises interesting phenomenology.

With this motivation, we present the phenomenological analysis of 3-4-1 model without exotic charges with parameters a =1, b =-1, c = 2. This model is embedded in an extended Pati-Salam gauge symmetry group $SU(3)_C \otimes SU(4)_L \otimes U(1)_{B-L} \otimes U(1)_{Y_0}$.

This paper is organized as follows. In section 2, we present the general formulation of the model. Section 3 deals with the scalar sector of the model and the physical Higgs mass spectrum. In section 4 we analyze the fermion mass spectrum. In section 5 we study the gauge boson sector and in section 6 the new neutral and charged currents in the model. Finally, section 7 deals with the results and conclusions.

**II General Formulation**

The symmetry breaking is achieved by introducing the scalars

$$\chi \sim \left(1,4,-\frac{1}{2}\right), \phi \sim \left(1,4,\frac{1}{2}\right), \phi_0 \sim (1,1,0).$$

We consider the decomposition

$$SU(4)_L \xrightarrow{\langle\phi\rangle} SU(3)_L \otimes U(1)_{Y_{1N}},$$

where $Y_{1N} = \dfrac{2}{\sqrt{6}} T_{15L} = Dg\left(\dfrac{1}{6}, \dfrac{1}{6}, \dfrac{1}{6}, -\dfrac{1}{2}\right)$ (5)

$$SU(3)_L \xrightarrow{\langle\chi\rangle} SU(2)_L \otimes U(1)_{X'} \text{ with } X' = \sqrt{3} T_{8L} = Dg\left(\dfrac{1}{2}, \dfrac{1}{2}, -1, 0\right) \quad (6)$$

$$U(1)_{Y_{1N}} \otimes U(1)_X \xrightarrow{\langle\phi_0\rangle} U(1)_{X_N} \text{ where } X_N = Y_{1N} + XI_4, X = \dfrac{(B-L)'}{2}. \quad (7)$$

The symmetry-breaking pattern for the gauge group is

$$SU(4)_L \otimes U(1)_X \xrightarrow{\langle\phi\rangle} SU(3)_L \otimes U(1)_{Y_{1N}} \otimes U(1)_X$$

$$\xrightarrow{\langle\chi\rangle} SU(2)_L \otimes U(1)_{X'} \otimes U(1)_{Y_{1N}} \otimes U(1)_X$$

$$\xrightarrow{\langle\phi_0\rangle} SU(2)_L \otimes U(1)_{X'} \otimes U(1)_{X_N} \xrightarrow{\langle\chi\rangle} SU(2)_L \otimes U(1)_Y$$

$$\xrightarrow{\langle\phi\rangle,\langle\chi\rangle} U(1)_Q$$

(8)

The hypercharge operator $Y = 2\left(-\dfrac{X'}{3} + X_N\right)$

The electric charge operator $Q = T_{3L} + \dfrac{Y}{2} = T_{3L} - \dfrac{1}{3} X' + Y_{1N} + XI_4$ (9)

We consider the 4-plet, $\overline{4}$-plet representations of $SU(4)_L$ with $SU(2)_L \otimes U(1)_{X'} \otimes U(1)_{Y_{1N}}$ content as

$$4 = \left(2_L, \dfrac{1}{2}, \dfrac{1}{6}\right) + \left(1_L, -1, \dfrac{1}{6}\right) + \left(1_L, 0, -\dfrac{1}{2}\right)$$

$$\overline{4} = \left(\overline{2}_L, -\dfrac{1}{2}, -\dfrac{1}{6}\right) + \left(1_L, +1, -\dfrac{1}{6}\right) + \left(1_L, 0, \dfrac{1}{2}\right)$$

The electric charges from eqn.(9)

$$Q[4] = Dg\left(\frac{1}{2}, -\frac{1}{2}, \frac{1}{2}, -\frac{1}{2}\right); Q[\bar{4}] = Dg\left(-\frac{1}{2}, \frac{1}{2}, -\frac{1}{2}, \frac{1}{2}\right).$$

**IIA. Gauge bosons:**

There are 24 gauge bosons for 3-4-1 gauge symmetry which include 8 gluons, a singlet and 15 gauge bosons for $SU(4)_L$. From eqn.(2), we note that $X = \frac{(B-L)'}{2}$ charge so that the 15 gauge bosons associated with $SU(4)_L$ do not carry $(B-L)$ and $Y_0$ charge and are not bilepton gauge bosons. Since we consider b = -1, c = 2, these include

$$\frac{1}{2}\lambda_\alpha A_\mu^\alpha = \frac{1}{\sqrt{2}}\begin{pmatrix} D_{1\mu}^0 & W_\mu^+ & K_\mu^0 & X_\mu^+ \\ W_\mu^- & D_{2\mu}^0 & K_{1\mu}^- & V_\mu^0 \\ K_\mu^{0*} & K_{1\mu}^+ & D_{3\mu}^0 & Y_\mu^+ \\ X_\mu^- & V_\mu^{0*} & Y_\mu^- & D_{4\mu}^0 \end{pmatrix} \qquad (10)$$

This is different from the case of b = 1, c = -2 as considered in [14], but corresponds to an 3-4-1 model which acts as a covering group for 3-3-1 Long model without bilepton gauge bosons..

**II B. Fermions:**

The anomaly-free models as suggested in [17] include three-family models with complete sets of spin $\frac{1}{2}$ Weyl spinors for which the 3-4-1 contents are

$$S_1^q = \left\{(t, b, U, D)_L \sim \left(3, 4, \frac{1}{6}\right), t_R^c \sim \left(\bar{3}, 1, -\frac{2}{3}\right), b_R^c \sim \left(\bar{3}, 1, \frac{1}{3}\right), U_R^c \sim \left(\bar{3}, 1, -\frac{2}{3}\right), D_R^c \sim \left(\bar{3}, 1, \frac{1}{3}\right)\right\}$$

$$S_2^q = \left\{(d_i, -u_i, D_i, U_i)_L, i = 1, 2 \sim \left(3, \bar{4}, \frac{1}{6}\right), d_{Ri}^c \sim \left(\bar{3}, 1, \frac{1}{3}\right), u^c{}_{Ri} \sim \left(\bar{3}, 1, -\frac{2}{3}\right), D_{Ri}^c \sim \left(\bar{3}, 1, \frac{1}{3}\right),\right.$$

$$\left. U_{Ri}^c \sim \left(\bar{3}, 1, -\frac{2}{3}\right)\right\}$$

$$S_3^l = \left\{(\nu_l, l^-, E^-, N^0)_L \sim \left(1, 4, -\frac{1}{2}\right), l_R^{c+} \sim (1,1,1), E_R^{c+} \sim (1,1,1)\right\}$$

- Model **F** : $S_1^q + 2 S_2^q + 3 S_3^l$

The phenomenology of this model has not been studied earlier. Model **E** has been studied in ref. [14] .We consider the following important difference from the 3-3-1 model [17,18,20,22]

- Addition of a new flavor $Y_0$ in 3-4-1 model where $X = Y_0 + \dfrac{(B-L)}{2}$. This flavor is considered for both Higgs scalars and right-handed singlet fermions.

- For left-right symmetric model, the electric charge operator Q satisfies [16]

$$Q = T_{3l} + T_{3R} + \dfrac{(B-L)}{2} I_4 ; T_{3R} = \sqrt{3} T_{8L} - \sqrt{6} T_{15L} = Dg\left(0, 0, \dfrac{1}{2}, -\dfrac{1}{2}\right)$$

The 3-3-1 gauge symmetry [17,18,20,22] can be obtained with the electric charge operator

$$Q = T_{3L} - \dfrac{1}{\sqrt{3}} T_{8L} + \left(\dfrac{4}{\sqrt{3}} T_{8L} - \sqrt{6} T_{15L} + \dfrac{(B-L)}{2} I_4\right) = T_{3L} - \dfrac{1}{\sqrt{3}} T_{8L} + X_1$$

The $X_1$ operator with additional conserved lepton charge $\dfrac{4}{\sqrt{3}} T_{8L}$ corresponds to the Long model. This fails for 3-4-1 case since $Q = X_1$ does not allow electric charges for $SU(4)_L$ singlet fermions.

The electric charge Q [4, X] is determined from eqn. (10) with addition of X charges. This gives non-exotic electric charges for quarks, leptons and Higgses.

$$Q[4, X] = Dg\left(\dfrac{1}{2} + X, -\dfrac{1}{2} + X, \dfrac{1}{2} + X, -\dfrac{1}{2} + X\right);$$

$$Q[\bar{4}, X] = Dg\left(-\dfrac{1}{2} + X, \dfrac{1}{2} + X, -\dfrac{1}{2} + X, \dfrac{1}{2} + X\right)$$

(12)

In the present model we can obtain bifundamental structures for matter fields as in 3-3-1 case [16]

The group is 3-4-1-1 with four basic fundamentals

$$G_C = SU(3)_C \otimes U(1)_{B-L} \otimes U(1)_{Y_0} : M = \left(3_C, \frac{1}{6}, 0\right), N = \left(1_C, -\frac{1}{2}, 0\right), N' = \left(1_C, 0, \frac{1}{2}\right).$$

$$G_L = SU(4)_L : a = 4_L.$$

The X charge is defined by eqn.(2) in this case.

Table **I** lists three-generation, anomaly-free fermions and Higgs scalars.

**TABLE I.** Three generations of anomaly-free fermions as bi-fundamentals of $3_C - 4_L - 1_{Y_0} - 1_{(B-L)}$.

| Fermions | Content | 3-4-1-1 | 3-4-1 |
|---|---|---|---|
| $Q_i, i = 1,2$ | $(d_i, -u_i, D_i, U_i)_L$ | $a^* M$ | $\left(3, \bar{4}, \frac{1}{6}\right)$ |
| $Q_3$ | $(t, b, U, D)_L$ | $aM$ | $\left(3, 4, \frac{1}{6}\right)$ |
| $d^c_{Ri}, i = 1,2$ | $(d^C_R, s^C_R)$ | $N'M^*$ | $\left(\bar{3}, 1, \frac{1}{3}\right)$ |
| $u^c_{Ri}, i = 1,2$ | $(u^C_R, c^C_R)$ | $N'^* M^*$ | $\left(\bar{3}, 1, -\frac{2}{3}\right)$ |
| $D^C_{iR}, i = 1,2$ | $(D^C_{iR}, D^C_{2R})$ | $N'M^*$ | $\left(\bar{3}, 1, \frac{1}{3}\right)$ |
| $U^C_{iR}, i = 1,2$ | $(U^C_{iR}, U^C_{2R})$ | $N'^* M^*$ | $\left(\bar{3}, 1, -\frac{2}{3}\right)$ |
| $t^C_R$ | | $N'^* M^*$ | $\left(\bar{3}, 1, -\frac{2}{3}\right)$ |
| $b_R^C$ | | $N'M^*$ | $\left(\bar{3},1,\frac{1}{3}\right)$ |
| $U_R^C$ | | $N'^*M^*$ | $\left(\bar{3},1,-\frac{2}{3}\right)$ |
| $D_R^C$ | | $N'M^*$ | $\left(\bar{3},1,\frac{1}{3}\right)$ |
| $\psi_{L\beta}, \beta=1,2,3$ | $\left(\nu_l, l^-, E_\beta^-, N_\beta^0\right)_L$ | $aN$ | $\left(1,4,-\frac{1}{2}\right)$ |
| $l_{R\beta}^C$ | $\left(e_R^C, \mu_R^C, \tau_R^C\right)$ | $N'N^*$ | $(1,1,1)$ |
| $E_{R\beta}^C$ | $\left(E_R^C, M_R^C, T_R^C\right)$ | $N'N^*$ | $(1,1,1)$ |

Scalars include two Higgs 4-plets

| | | | |
|---|---|---|---|
| $\chi$ | $\left(\chi_1^0, \chi_2^-, \chi_3^0, \chi_4^-\right)$ | $aN'^*$ | $\left(1,4,-\frac{1}{2}\right)$ |
| $\phi$ | $\left(\phi_1^+, \phi_2^0, \phi_3^+, \phi_4^0\right)$ | $aN'$ | $\left(1,4,\frac{1}{2}\right)$ |

Additional singlet scalar

| | | | |
|---|---|---|---|
| $\phi_0$ | | $NN'NN'$ | $(1,1,0)$ |

_______________________________________________________________

–

## II C Scalars:

We assume that only two scalars can give masses to all fermions as suggested for 3-3-1 case in ref [ 18,20] .These are

$$\chi \sim \left[1,4,-\frac{1}{2}\right] = \begin{pmatrix} \chi_1^0 \\ \chi_2^- \\ \chi_3^0 \\ \chi_4^- \end{pmatrix}, \text{ with } Y_0 = -\frac{1}{2}; \phi \sim \left(1,4,\frac{1}{2}\right) = \begin{pmatrix} \phi_1^+ \\ \phi_2^0 \\ \phi_3^+ \\ \phi_4^0 \end{pmatrix}, Y_0 = \frac{1}{2} \qquad (13)$$

These scalars are introduced with Vacuum Expectation Values (VEV) aligned in the direction $\langle \phi \rangle = \left(0, \frac{v}{\sqrt{2}}, 0, \frac{z}{\sqrt{2}}\right)^T, \langle \chi \rangle = \left(\frac{u}{\sqrt{2}}, 0, \frac{V}{\sqrt{2}}, 0\right)^T$ (14)

We introduce an additional singlet scalar $\phi_0 \sim [1,1,0]$ with L = -2 and zero X charge.

$$\phi_0 : X = (B-L)' = 0 = \frac{(B-L)}{2} + Y_0; \frac{(B-L)}{2} = 1, Y_0 = -1.$$

The VEV $\langle \phi_0 \rangle = w.$ This breaks the symmetry as $U(1)_{B-L} \otimes U(1)_{Y_0} \xrightarrow{\langle \phi_0 \rangle} U(1)_{(B-L)'}$

## III. Scalar sector and Higgs mass spectrum

The most general Higgs potential with two Higgs 4-plets is given by

$$V(\chi,\varphi) = \mu_1^2 \chi^\dagger \chi + \mu_2^2 \varphi^\dagger \varphi + \lambda_1 \left(\chi^\dagger \chi\right)^2 + \lambda_2 \left(\varphi^\dagger \varphi\right)^2 + \lambda_3 \left(\chi^\dagger \chi\right)\left(\varphi^\dagger \varphi\right) + \lambda_4 \left(\chi^\dagger \varphi\right)\left(\varphi^\dagger \chi\right)$$

With the addition of a neutral singlet scalar $\phi_0$, the total potential is

$$V_T(\varphi,\chi,\phi_0) = V(\chi,\phi) + \mu_\phi^2 \phi_0^\dagger \phi_0 + \lambda_0 \left(\phi_0^\dagger \phi_0\right)^2 + \phi_0^\dagger \phi_0 \left[\lambda_\varphi \varphi^\dagger \varphi + \lambda_\chi \chi^\dagger \chi\right] \qquad (15)$$

where $\mu's$ have mass dimensions while $\lambda's$ are dimensionless

The neutral Higgses are shifted according to

$$\chi_1^0 = \frac{1}{\sqrt{2}}(u + S_1 + iA_1); \chi_3^0 = \frac{1}{\sqrt{2}}(V + S_3 + iA_3); \varphi_2^0 = \frac{1}{\sqrt{2}}(v + S_2 + iA_2);$$

$$\varphi_4^0 = \frac{1}{\sqrt{2}}(z + S_4 + iA_4); \phi_0 = \frac{1}{\sqrt{2}}(w + S_0) \qquad (16)$$

The VEV's u, v << V, w, z is assumed for these scalars.

### IIIA. The Constraint Equations

Due to the requirement of minimum potential at the chosen VEV's, we consider

$$\frac{\partial V_T}{\partial u} = 0 = \frac{\partial V_T}{\partial v} = \frac{\partial V_T}{\partial V} = \frac{\partial V_T}{\partial z} = \frac{\partial V_T}{\partial w} \qquad (17)$$

The constraint equations are

$$\mu_1^2 + \lambda_1\left(u^2 + V^2\right) + \lambda_3\frac{(v^2 + z^2)}{2} + \lambda_\chi\frac{w^2}{2} = 0$$

$$\mu_2^2 + \lambda_2\left(v^2 + z^2\right) + \frac{\lambda_3}{2}\left(u^2 + V^2\right) + \frac{\lambda_\phi}{2}w^2 = 0$$

$$\mu_0^2 + \lambda_\chi\frac{\left(u^2 + V^2\right)}{2} + \lambda_\phi\frac{\left(v^2 + z^2\right)}{2} + \lambda_0 w^2 = 0 \qquad (18)$$

Imposing the constraint equations we get the minimum value and mass terms:

$$V_{\min} = -\frac{\lambda_2}{4}\left(v^2 + z^2\right)^2 - \frac{1}{4}(u^2 + V^2)\left[\lambda_1(u^2 + V^2) + \lambda_3\left(v^2 + z^2\right)\right] - \frac{\lambda_0}{4}w^4$$
$$-\frac{1}{4}w^2\left[\lambda_\chi(u^2 + V^2) + \lambda_\phi\left(v^2 + z^2\right)\right] \qquad (19)$$

This includes extra terms due to the additional scalar $\phi_0$ which has $B - L = 2$.

### III B: CP-even and CP-odd neutral Higgs mass spectrum

In the basis $\begin{pmatrix} S_1 & S_3 & S_2 & S_4 & S_0 \end{pmatrix}$, the mass-squared mass matrix is given by

$$M^2{}_H = \frac{\partial^2 V_T}{\partial S_i \partial S_j}, \text{ where } i, j = 1,2,3,4,0.$$

The 5 x 5 mass-squared matrix for neutral scalars is

$$M_N{}^2 = \begin{pmatrix} 2\lambda_1 u^2 & \lambda_1 uV & \lambda_3 uv & \lambda_3 uz & \lambda_\chi uw \\ \lambda_1 uV & 2\lambda_1 V^2 & \lambda_3 vV & \lambda_3 Vz & \lambda_\chi Vw \\ \lambda_3 uv & \lambda_3 vV & 2\lambda_2 v^2 & \lambda_2 vz & \lambda_\phi zw \\ \lambda_3 uz & \lambda_3 vz & \lambda_2 vz & 2\lambda_2 z^2 & \lambda_\phi vw \\ \lambda_\chi uw & \lambda_\chi Vw & \lambda_\phi zw & \lambda_\phi vw & 2\lambda_0 w^2 \end{pmatrix} \qquad (20)$$

Two Goldstone bosons and three massive Higgses $\left(G_1^0, G_2^0, H_1^0, H_2^0, H_3^0\right)$ are obtained

The mass of $H_3^0$, $m_{H_3^0}^2 = 2\lambda_0 w^2$ is obtained in the approximation $\lambda_\phi \ll \lambda_0$

The Goldstone boson $G_1^0$ and Higgs $H_1^0$ are given in terms of scalars $S_1, S_3$ by

$$G_1^0 = \cos\theta S_1 - \sin\theta S_3; H_1^0 = \sin\theta S_1 + \cos\theta S_3 \tag{21}$$

where $m_{G_1^0}^2 = 0, m_{H_1^0}^2 = 2\lambda_1(u^2 + V^2)$ (22)

$$\tan 2\theta = \frac{uV}{(u^2 - V^2)}$$

The Goldstone boson $G_2^0$ and Higgs $H_2^0$ are given in terms of scalars $S_2, S_4$ by

$$G_2^0 = \cos\alpha S_2 - \sin\alpha S_4; H_2^0 = \sin\alpha S_2 + \cos\alpha S_4 \tag{23}$$

where $m_{G_2^0}^2 = 0, m_{H_2^0}^2 = 2\lambda_2(v^2 + z^2)$ (24)

$$\tan 2\alpha = \frac{vz}{(v^2 - z^2)}$$

The SM Higgs given by $H_2^0$ is heavy due to VEV z.

The pseudoscalars are given by four Goldstone bosons $G_3^0, G_4^0, G_5^0, G_6^0$ which satisfy

$$G_3^0 = A_3, G_4^0 = A_4, G_5^0 = A_5, G_6^0 = A_6 \tag{25}$$

**IIIC: Charged Higgs mass spectrum**

In the basis $(\phi_1^+ \ \phi_3^+ \ \chi_2^+ \ \chi_4^+)$ the mass squared matrix is obtained from the scalar potential for charged Higgses. The physical Higgses include three Goldstone bosons and one massive single-charged scalar physical Higgs.

$$M^2_{charged} = \begin{pmatrix} u^2 & uV & uv & uz \\ uV & V^2 & vV & Vz \\ uv & vV & v^2 & vz \\ uz & Vz & vz & z^2 \end{pmatrix} \tag{26}$$

The three charged Goldstone bosons $G_7^+, G_8^+, G_9^+$ and one massive physical scalar $H^+$ are obtained with squared mass.

$$m^2_{H^+} = \frac{\lambda_3}{4}\left(u^2 + v^2 + V^2 + z^2\right), m^2_{G_7} = 0 = m^2_{G_8} = m^2_{G_9} \tag{27}$$

The main results for the Higgs scalars are

- There are nine Goldstone bosons, three massive neutral and one charged physical scalar fields

- The masses of scalars are

$$. m^2_{H_1^0} = 2\lambda_1\left(u^2 + V^2\right); m^2_{H_2^0} = 2\lambda_2\left(v^2 + z^2\right); m^2_{H_3^0} = 2\lambda_0 w^2;$$

$$m^2_{H^+} = \frac{\lambda_3}{4}\left(u^2 + v^2 + V^2 + z^2\right)$$

- There is no CP-violation in the scalar sector due to absence of massive pseudoscalar field.

### IV. Fermion masses

The Yukawa interactions for quarks are now obtained for three generations

where $\alpha = 1, 2$; $i = 1, 2, 3$ are generation indices

$$L_Y^Q = h^u_{3i}\left(3, 4, \frac{1}{6}\right)x\left(1, \bar{4}, \frac{1}{2}\right)x\left(\bar{3}, 1, -\frac{2}{3}\right) + h^d_{3i}\left(3, 4, \frac{1}{6}\right)x\left(1, \bar{4}, -\frac{1}{2}\right)x\left(\bar{3}, 1, \frac{1}{3}\right)$$

$$+ h^d_{\alpha i}\left(3, \bar{4}, \frac{1}{6}\right)x\left(1, 4, -\frac{1}{2}\right)x\left(\bar{3}, 1, \frac{1}{3}\right) + h^u_{\alpha i}\left(3, \bar{4}, \frac{1}{6}\right)x\left(1, 4, \frac{1}{2}\right)x\left(\bar{3}, 1, -\frac{2}{3}\right) \tag{28}$$

$$L_Y^Q = \sum_i (h^t_{3i}\bar{t}_L\chi_1^0 u_{iR} + h^U_{3i}\bar{U}_L\chi_3^0 u_{iR} + h^b_{3i}\bar{b}_L\phi_2^0 d_{iR} + h^D_{3i}\bar{D}_L\phi_4^0 d_{iR}) +$$

$$\sum_{\alpha,i}\left(h^d_{\alpha i}\bar{d}_{\alpha L}\chi_1^{0*} d_{iR} + h^u_{\alpha i}\bar{u}_{\alpha L}\phi_2^{0*} u_{iR} + h^D_{\alpha i}\bar{D}_{\alpha L}\chi_3^{0*} d_{iR} + h^U_{\alpha i}\bar{U}_{\alpha L}\phi_4^{0*} u_{iR}\right) +$$

$$\bar{t}_L\chi_1^0(h^t_{34}U_R + h'^t_3 U_{1R} + h''^t_3 U_{2R}) + \bar{U}_L\chi_3^0(h^U_{34}U_R + h'^U_3 U_{1R} + h''^U_3 U_{2R}) +$$

$$\bar{b}_L\phi_2^0(h^D_{34}D_R + h'^b_3 D_{1R} + h''^b_3 D_{2R}) + \bar{D}_L\phi_4^0(h^D_{34}D_R + h^D_1 D_{1R} + h^D_2 D_{2R}) +$$

$$\sum_\alpha [\bar{d}_{\alpha L}\chi_1^{0*}(h_{14}^d D_R + h'_\alpha D_{1R} + h''_\alpha D_{2R}) + \bar{u}_{\alpha L}\phi_2^{0*}(h_{14}^u U_R + h'_\alpha U_{1R} + h''_\alpha U_{2R}) +$$

$$\bar{D}_{\alpha L}\chi_3^{0*}(h_{\alpha 4}^{D_\alpha} D_R + h_1^{D_\alpha} D_{1R} + h_2^{D_\alpha} D_{2R}) + \bar{U}_{\alpha L}\phi_4^{0*}(h_{\alpha 4}^{U_\alpha} U_R + h_1^{U_\alpha} U_{1R} + h_2^{U_\alpha} U_{2R})] + h.c. \quad (29)$$

Both the Higgs scalars and right-handed fermions are assigned flavor $Y_0 = \pm\frac{1}{2}$

The lepton Yukawa terms are given for three-generations of leptons. The singlet fermions do not play any role in anomaly cancellations. We consider the leptons in eqn.(11)

$$S_3^l = \psi_L\left(1,4,-\frac{1}{2}\right), l^-_R(1,1,-1), E^-_R(1,1,-1). \quad (30)$$

where $\psi_L\left(1,4,-\frac{1}{2}\right) \sim \left(v_l, l^-, N^0, E^-\right)_L$

The Yukawa Lagrangian for charged leptons

$$L_Y^l = \sum_{ij}(h_{ij}^l \bar{\psi}_{Li}\phi l_{Rj} + h_{ij}^E \bar{\psi}_{Li}\phi E_{Rj}) + h.c. \quad (31)$$

here i, j = 1,2,3.

## IVA. The quark sector

From the Yukawa Lagrangian for quarks we obtain the tree level mass matrix for up- and down quarks. For up-type quarks in the basis $(u_1 \ u_2 \ u_3 \ U \ U_1 \ U_2)$

$$M_u = -\frac{1}{\sqrt{2}}\begin{pmatrix} h_{11}^u v & h_{12}^u v & h_{13}^u v & h_{14}^u v & h'_1 v & h''_1 v \\ h_{21}^u v & h_{22}^u v & h_{23}^u v & h_{24}^u v & h'_2 v & h''_2 v \\ -h_{31}^t u & -h_{32}^t u & -h_{33}^t u & -h_{34}^t u & -h'_3 u & -h''_3 u \\ -h_{31}^U V & -h_{32}^U V & -h_{33}^U V & -h_{34}^U V & -h_3^U V & -h_3^U V \\ h_{11}^{U_1} z & h_{12}^{U_1} z & h_{13}^{U_1} z & h_{14}^{U_1} z & h_1^{U_1} z & h_2^{U_1} z \\ h_{11}^{U_2} z & h_{12}^{U_2} z & h_{13}^{U_2} z & h_{14}^{U_2} z & h_1^{U_2} z & h_2^{U_2} z \end{pmatrix} \quad (32)$$

If we use the SM assumption and let only the flavor diagonal couplings be finite[17], we obtain two heavy up-type quarks $U_1, U_2$ with masses proportional to VEV z, where $z \gg u, v, V$

$$M_{U_i} = -\frac{z}{\sqrt{2}}\left(h_1^{U_1} + h_2^{U_2}\right), i = 1,2. \quad (33)$$

There is an additional up quark U in the third generation along with top quark,

$$M_U = h^U_{34} \frac{V}{\sqrt{2}}, \quad M_t = h^t_{33} \frac{u}{\sqrt{2}}, \tag{34}$$

This gives two massive top quarks for mixing between t and U states.

The up quark and charm quark masses are

$$M_u = -h^u_{11} \frac{v}{\sqrt{2}}, \quad M_c = -h^u_{22} \frac{v}{\sqrt{2}} \tag{35}$$

The mixing of u, c with $U_1, U_2$ gives one massless and one massive quark each for generation 1,2. The exotic quarks include $U', C'$

$$U' = \frac{vu_1 + zU_1}{\sqrt{v^2 + z^2}}, \quad C' = \frac{vu_2 + zU_2}{\sqrt{v^2 + z^2}}$$

For the down-type quarks in the basis $(d_1 \quad d_2 \quad d_3 \quad D \quad D_1 \quad D_2)$ the tree-level mass-matrix

$$M_d = -\frac{1}{\sqrt{2}} \begin{pmatrix} h^d_{11}u & h^d_{12}u & h^d_{13}u & h^d_{14}u & h'_1 u & h''_1 u \\ h^d_{21}u & h^d_{22}u & h^d_{23}u & h^d_{24}u & h'_2 u & h''_2 u \\ -h^b_{31}v & -h^b_{32}v & -h^b_{33}v & -h^b_{34}v & -h'_3 v & -h''_3 v \\ h^D_{31}z & h^D_{32}z & h^D_{33}z & h^D_{34}z & h^D_1 z & h^D_2 z \\ -h^{D_1}_{11}V & -h^{D_1}_{12}V & -h^{D_1}_{13}V & -h^{D_1}_{14}V & -h^{D_1}_1 V & -h^{D_1}_2 V \\ -h^{D_2}_{21}V & -h^{D_2}_{22}V & -h^{D_2}_{23}V & -h^{D_2}_{24}V & -h^{D_2}_1 V & -h^{D_2}_2 V \end{pmatrix} \tag{36}$$

We use the SM assumption and let only the flavor diagonal couplings be finite. There is an additional down-quark D in the third generation along with bottom quark,

$$M_D = h^D_{34} \frac{z}{\sqrt{2}}, \quad M_b = h^b_{33} \frac{v}{\sqrt{2}}. \tag{37}$$

while the down quark and strange quark masses are

$$M_d = -h^d_{11} \frac{u}{\sqrt{2}}, \quad M_s = -h^d_{22} \frac{u}{\sqrt{2}} \tag{38}$$

Due to mixing of down and strange quarks with exotic quarks ( d,D$_1$),(s,D$_2$) one obtains one massless and one heavy quark in each case. The massive down quarks are $D' = \dfrac{ud_1 + VD_1}{\sqrt{u^2 + V^2}}$,

$$S' = \dfrac{ud_2 + VD_2}{\sqrt{u^2 + V^2}}$$

The ordinary quarks u, d, c, s require radiative corrections for masses as suggested recently [15]

**IVB: The lepton sector**

From the lepton Yukawa terms, the mass of charged leptons is obtained with diagonal coupling constants $h^l_{ij}$, i,j = 1,2,3 as in SM

$$m^e = \dfrac{h^e_{11}}{\sqrt{2}} v, \quad m^\mu = \dfrac{h^\mu_{22}}{\sqrt{2}} v, \quad m^\tau = \dfrac{h^\tau_{33}}{\sqrt{2}} v \tag{39}$$

The charged leptons E,M,T

$$m^E = \dfrac{h^E_{11}}{\sqrt{2}} z, \quad m^M = \dfrac{h^M_{22}}{\sqrt{2}} z, \quad m^T = \dfrac{h^T_{33}}{\sqrt{2}} z \tag{40}$$

For neutrinos, we obtain massless, neutral states $(\nu_1 \ \nu_2 \ \nu_3 \ N_1 \ N_2 \ N_3)$ without adding extra singlet right-handed neutrinos. The analysis of the lepton sector is similar to the b =1,c = -2 case [14] which has the leptons in $S^l_4$ instead of $S^l_3$ of eqn.(11)

$$S^l_4 = \psi_L \left(1, \overline{4}, -\dfrac{1}{2}\right) \square \left(e^-, \nu, N^0, E^-\right), l^+{}_R (1,1,+1), E^+{}_R (1,1,+1). \tag{41}$$

Additional lepton-number violating interactions can be introduced as $\psi_L \psi_L \overline{\chi\chi} \phi_0$ to give neutrino masses in a 3-4-1 model.

**V. Gauge boson sector**

The case of b = -1, c = 2 in the electric charge operator gives the 15-plet gauge bosons of SU(4)$_L$ as in eqn.( 10).The covariant derivative for 4-plet is

$$D^\mu = (\partial^\mu - ig T_\alpha W^\mu_\alpha - ig_X X I_4 B^\mu), \text{ where } \alpha = 1, 2, ..15; \mu = 1, 2, 3, 4. \tag{42}$$

The coupling constants of SU(4)$_L$ and U(1)$_X$ are g and g$_X$ respectively

The neutral components in eqn.(10) are given by

$$D_1^{\mu 0} = \frac{W_3^\mu}{\sqrt{2}} + \frac{W_8^\mu}{\sqrt{6}} + \frac{W_{15}^\mu}{\sqrt{12}}; D_2^{\mu 0} = -\frac{W_3^\mu}{\sqrt{2}} + \frac{W_8^\mu}{\sqrt{6}} + \frac{W_{15}^\mu}{\sqrt{12}}; D_3^{\mu 0} = -2\frac{W_8^\mu}{\sqrt{6}} + \frac{W_{15}^\mu}{\sqrt{12}}; D_4^{\mu 0} = -3\frac{W_{15}^\mu}{\sqrt{12}} \quad (43)$$

we define $D_\mu \equiv \partial_\mu - iP_\mu$ where

$$P_\mu =$$

$$\frac{g}{2}\begin{pmatrix} W_{3\mu} + \frac{1}{\sqrt{3}}W_{8\mu} + \frac{1}{\sqrt{6}}W_{15\mu} + 2tXB_\mu & \sqrt{2}W_\mu'^+ & \sqrt{2}K_\mu'^0 & \sqrt{2}X_\mu'^+ \\ \sqrt{2}W_\mu'^- & -W_{3\mu} + \frac{1}{\sqrt{3}}W_{8\mu} + \frac{1}{\sqrt{6}}W_{15\mu} + 2tXB_\mu & \sqrt{2}K_{1\mu}'^- & \sqrt{2}V_\mu'^0 \\ \sqrt{2}K_\mu'^{0*} & \sqrt{2}K_{1\mu}'^+ & -\frac{2}{\sqrt{3}}W_{8\mu} + \frac{1}{\sqrt{6}}W_{15\mu} + 2tXB_\mu & \sqrt{2}Y_\mu'^+ \\ \sqrt{2}X_\mu'^- & \sqrt{2}V_\mu'^{0*} & \sqrt{2}Y_\mu'^- & -\frac{3}{\sqrt{6}}W_{15\mu} + 2tXB_\mu \end{pmatrix}$$

Here $t = \frac{g_X}{g}$. \hfill (44)

We consider the combinations

$$W'^\pm \equiv \frac{W_1 \mp iW_2}{\sqrt{2}}; K'^0 \equiv \frac{W_4 - iW_5}{\sqrt{2}}; K^\mp_1 \equiv \frac{W_6 \mp iW_7}{\sqrt{2}}$$

$$X'^\mp = \frac{W_9 \mp iW_{10}}{\sqrt{2}}, V'^0 = \frac{W_{11} - iW_{12}}{\sqrt{2}}, Y'^\pm = \frac{W_{13} \mp iW_{14}}{\sqrt{2}} \quad (45)$$

The covariant derivative for 4-plets is now used for obtaining gauge boson masses using

$\chi\left[1,4,-\frac{1}{2}\right], \phi\left[1,4,\frac{1}{2}\right]$, 4-plet Higgs scalars

$$L_{GB} = \left(D_\mu \langle \phi \rangle\right)^\dagger \left(D^\mu \langle \phi \rangle\right) + \left(D_\mu \langle \chi \rangle\right)^\dagger \left(D^\mu \langle \chi \rangle\right)$$

$$= \frac{g^2}{4}\left(u^2 + v^2\right)W_\mu'^{-}W'^{+\mu} + \frac{g^2}{4}\left(V^2 + v^2\right)K_{1\mu}'^{-}K_1'^{+\mu} + \frac{g^2}{4}\left(u^2 + z^2\right)X_\mu'^{-}X'^{+\mu}$$

$$+ \frac{g^2}{4}\left(V^2 + z^2\right)Y_\mu'^{-}Y'^{+\mu} + \frac{g^2}{4}\left(W_\mu'^{-}K_1'^{+\mu} + K_{1\mu}'^{-}W'^{+\mu}\right)uV + \frac{g^2}{4}\left(Y_\mu'^{-}K_1'^{+\mu} + K_{1\mu}'^{-}Y'^{+\mu}\right)vz$$

$$+ \frac{g^2}{4}\left(Y_\mu'^{+}X'^{-\mu} + X_\mu'^{+}Y'^{-\mu}\right)uV + \frac{g^2 u^2}{8}\left(W_{3\mu} + \frac{1}{\sqrt{3}}W_{8\mu} + \frac{1}{\sqrt{6}}W_{15\mu} - tB_\mu\right)^2$$

$$+ \frac{g^2 v^2}{8}\left(-W_{3\mu} + \frac{1}{\sqrt{3}}W_{8\mu} + \frac{1}{\sqrt{6}}W_{15\mu} + tB_\mu\right)^2 + \frac{g^2 V^2}{8}\left(-\frac{2}{\sqrt{3}}W_{8\mu} + \frac{1}{\sqrt{6}}W_{15\mu} - tB_\mu\right)^2$$

$$+ \frac{g^2 z^2}{8}\left(-\frac{3}{\sqrt{6}}W_{15\mu} + tB_\mu\right)^2 + \frac{g^2}{16}\left(u^2 + V^2\right)\left[\left(K'^{0\mu} + K'^{*0\mu}\right)^2 + \left\{i\left(K'^{0\mu} - K'^{*0\mu}\right)\right\}^2\right]$$

$$+ \frac{g^2}{16}\left(v^2 + z^2\right)\left[\left(V'^{0\mu} + V'^{*0\mu}\right)^2 + \left\{i\left(V'^{0\mu} - V'^{*0\mu}\right)\right\}^2\right]$$

$$\frac{g^2 uV}{4\sqrt{2}}\left(W_{3\mu} + \frac{1}{\sqrt{3}}W_{8\mu} + \frac{1}{\sqrt{6}}W_{15\mu} - tB_\mu\right)\left(K'^{0\mu} + K'^{*0\mu}\right)$$

$$+ \frac{g^2 vz}{4\sqrt{2}}\left(-W_{3\mu} + \frac{1}{\sqrt{3}}W_{8\mu} + \frac{1}{\sqrt{6}}W_{15\mu} + tB_\mu\right)\left(V'^{0\mu} + V'^{*0\mu}\right)$$

$$+ \frac{g^2 uV}{4\sqrt{2}}\left(-\frac{2}{\sqrt{3}}W_{8\mu} + \frac{1}{\sqrt{6}}W_{15\mu} - tB_\mu\right)\left(K'^{0\mu} + K'^{*0\mu}\right) + \frac{g^2 vz}{4\sqrt{2}}\left(-\frac{3}{\sqrt{6}}W_{15\mu} + tB_\mu\right)\left(V'^{0*} + V'^{0}\right)$$

(46)

We consider $u \ll u, v, V$ for the VEV's.

## VA. Charged gauge bosons

An important point of difference from 3-4-1 models without exotic charges [14,15] is that for three and four scalar Higgses there is no mixing between SM gauge bosons and new charged gauge fields. However, in 3-3-1 two Higgs triplet model, there is mixing between charged gauge bosons $W_\mu'^{\pm}$ and $K_{1\mu}'^{\pm}$ {18, 20}.

The 4-plet scalars $\chi, \phi$ introduces mixing among SM gauge boson $W'^{\pm}_{\mu}$ and charged gauge bosons $K'^{\pm}_{1\mu}, X'^{\pm}_{\mu}$. We assume $u \ll v, V, z$, and neglect mixing of $W'^{\pm}_{\mu}$ with $X'^{\pm}_{\mu}$

The masses for additional charged gauge bosons are as follows:

$$M^2_Y = \frac{g^2}{4}(V^2 + z^2), \quad M^2_X = \frac{g^2}{4}(u^2 + z^2) \quad (47)$$

The SM gauge boson $W'^{\pm}$ and $K'^{\pm}_{1\mu}$ mix according to the matrix below in 3-4-1 case.

$$L^{charged}_{GB} = \frac{g^2}{4}\begin{pmatrix} W'^{-}_{\mu} & K'^{-}_{1\mu} \end{pmatrix}\begin{pmatrix} u^2+v^2 & uV \\ uV & v^2+V^2 \end{pmatrix}\begin{pmatrix} W'^{+\mu} \\ K'^{+\mu}_1 \end{pmatrix} \quad (48)$$

The physical charged gauge bosons $W^{\pm}_{\mu}, K^{\pm}_{1\mu}$ are

$$W^{-}_{\mu} = \cos\beta W'^{-}_{\mu} + \sin\beta K'^{-}_{1\mu}$$
$$K^{-}_{1\mu} = -\sin\beta W'^{-}_{\mu} + \cos\beta K'^{-}_{1\mu}$$

The mixing angle $\beta$ is given by $\tan\beta = \frac{u}{V}$ \quad (49)

For $\beta$ to be very small, $u \ll V$. In this case, $W^{\pm}_{\mu} = W'^{\pm}_{\mu}, K^{\pm}_{1\mu} = K'^{\pm}_{1\mu}$. The mass of physical gauge bosons are defined by eqn.(49)

$$M^2_W = \frac{g^2 v^2}{4}; M^2_{K_1} = \frac{g^2}{4}(u^2 + v^2 + V^2) \quad (50)$$

This gives $v = v_{weak} = 246$ GeV as in 3-3-1 case.

**V B. Neutral gauge boson sector**

Due to two 4-plet scalars there is mixing between the SM neutral gauge bosons and new neutral gauge bosons $K'^0_{\mu}, V'^0_{\mu}$. This is similar to the 3-3-1 case where SM gauge bosons mix with $K'^0_{\mu}$. In the 3-4-1 case we consider $u \ll v, V, z$ so that $B'_{\mu}, W'_{3,8,15\mu}$ mix with only $K'^0_{\mu}$, $K'^{0*}_{\mu}$. The real and imaginary parts of $K'^0_{\mu}$ and $K'^{0*}_{\mu}$ are $W_{4\mu}, W_{5\mu}$ respectively where

$$W_{4\mu} = \frac{1}{\sqrt{2}}\left(K_{\mu}^{\prime 0} + K_{\mu}^{\prime 0*}\right), W_{5\mu} = \frac{i}{\sqrt{2}}\left(K_{\mu}^{\prime 0} - K_{\mu}^{\prime 0*}\right)$$

In the basis $\begin{pmatrix} W_3 & W_8 & W_{15} & B & W_4 & W_{11}\end{pmatrix}$, the mass matrix is given by

$$M_N^2 = \frac{g^2}{4}\begin{pmatrix} u^2+v^2 & \frac{1}{\sqrt{3}}(v^2-u^2) & \frac{1}{\sqrt{6}}(u^2-v^2) & \frac{t}{2}(-u^2+v^2) & 2uV & -2vz \\ \frac{1}{\sqrt{3}}(v^2-u^2) & \frac{1}{3}(u^2+v^2+4V^2) & \frac{1}{3\sqrt{2}}(u^2+v^2-2V^2) & \frac{t}{2\sqrt{3}}(u^2+v^2+4V^2) & -\frac{2}{\sqrt{3}}uV & \frac{2}{\sqrt{3}}vz \\ \frac{1}{\sqrt{6}}(u^2-v^2) & \frac{1}{3\sqrt{2}}(u^2+v^2-2V^2) & \frac{1}{6}(u^2+v^2+V^2+9z^2) & -\frac{t}{2\sqrt{6}}(u^2+v^2+V^2-3z^2) & \frac{4}{\sqrt{6}}uV & \frac{2}{\sqrt{6}}vz \\ \frac{t}{2}(-u^2+v^2) & -\frac{t}{2\sqrt{3}}(u^2+v^2+4V^2) & -\frac{t}{2\sqrt{6}}(u^2+v^2+V^2-3z^2) & \frac{t^2}{4}(u^2+v^2+V^2+z^2) & -2tuV & -tvz \\ 2uV & -\frac{2}{\sqrt{3}}uV & \frac{4}{\sqrt{6}}uV & -2tuV & u^2+V^2 & 0 \\ -2vz & \frac{2}{\sqrt{3}}vz & \frac{2}{\sqrt{6}}vz & -tvz & 0 & v^2+z^2 \end{pmatrix}$$

(51)

The mass for additional neutral gauge boson $V^0$ for $v \square z$ is

$$M_V^2 = \frac{g^2}{4}\left(v^2 + z^2\right) \tag{52}$$

For symmetry-breaking, we consider eqn.(8),

$$U(1)_{Y_{1N}} \otimes U(1)_X \xrightarrow{\langle \phi_0 \rangle} U(1)_{X_N} \; ; X_N = Y_{1N} + XI_4.$$

A new neutral gauge boson $V_{1\mu}$ is associated with $U(1)_X$. The physical gauge fields $V_{1\mu}$ and a fourth neutral gauge boson $Z''$ are defined as

$$Z''_\mu = -\sin\theta_N W_{15\mu} + \cos\theta_N B_\mu \; ; V_{1\mu} = \cos\theta_N W_{15\mu} + \sin\theta_N B_\mu \tag{53}$$

Since $\left|D_\mu \phi\right|^2 = \frac{g^2}{8} z^2 \left[-\frac{3}{\sqrt{6}}\left(-\sin\theta_N Z''_\mu + \cos\theta_N V_{1\mu}\right) + t\left(\cos\theta_N Z''_\mu + \sin\theta_N V_{1\mu}\right)\right]^2$

for $\tan\theta_N = \sqrt{\dfrac{3}{2}}\dfrac{g}{g_X} = \sqrt{\dfrac{3}{2}}\dfrac{1}{t}$, $Z''_\mu$ acquires a mass squared $M^2_{Z''} = \dfrac{g^2}{4}z^2\left(t^2 + \dfrac{3}{2}\right)$ while the mass of $V_{1\mu}$ is independent of VEV z. From eqn.(55),

$$V_{1\mu} = \dfrac{1}{\sqrt{t^2+\dfrac{3}{2}}}\left(tW_{15\mu} + \sqrt{\dfrac{3}{2}}B_\mu\right) \tag{54}$$

For mixing of $W_{8\mu}, V_{1\mu}$

$$\begin{aligned}Y_\mu &= -\cos\phi\, W_{8\mu} + \sin\phi\, V_{1\mu};\quad Z'_\mu = \sin\phi\, W_{8\mu} + \cos\phi\, V_{1\mu} \\ A_\mu &= \sin\theta_W W_{3\mu} + \cos\theta_W Y_\mu;\quad Z_\mu = \cos\theta_W W_{3\mu} - \sin\theta_W Y_\mu\end{aligned} \tag{55}$$

where $\cos\phi = \dfrac{\tan\theta_W}{\sqrt{3}},\ \sin\phi = \sqrt{1-\dfrac{\tan^2\theta_W}{3}}$

Substituting for $Y_\mu$,

$$\begin{aligned}A_\mu &= \sin\theta_W W_{3\mu} + \cos\theta_W\left(-\dfrac{\tan\theta_W}{\sqrt{3}}W_{8\mu} + \sqrt{1-\dfrac{\tan^2\theta_W}{3}}V_{1\mu}\right), \\ Z_\mu &= \cos\theta_W W_{3\mu} - \sin\theta_W\left(-\dfrac{\tan\theta_W}{\sqrt{3}}W_{8\mu} + \sqrt{1-\dfrac{\tan^2\theta_W}{3}}V_{1\mu}\right), \\ Z'_\mu &= \sqrt{1-\dfrac{\tan^2\theta_W}{3}}W_{8\mu} + \dfrac{\tan\theta_W}{\sqrt{3}}V_{1\mu}.\end{aligned} \tag{56}$$

This is identical to the 3-3-1 case with the new vector $V_{1\mu}$ replacing $B_\mu$ in the Long model[18]

The unification condition is

$$\dfrac{1}{g_Y^2} = \dfrac{1}{g^2} + \dfrac{1}{g_X^2};\quad \tan\theta_W = \dfrac{g_Y}{g} = \dfrac{g_X}{\sqrt{g_X^2+g^2}} \tag{57}$$

Substituting for $V_{1\mu}$, we obtain the photon field $A_\mu$, $Z_\mu, Z'_\mu$ as

$$A_\mu = \sin\theta_W W_{3\mu} + \cos\theta_W\left[\dfrac{-\tan\theta_W}{\sqrt{3}}\left(W_{8\mu} - \sqrt{2}W_{15\mu}\right) + \sqrt{1-\tan^2\theta_W}\,B_\mu\right]$$

$$Z_\mu = \cos\theta_W A_{3\mu} - \sin\theta_W \left[ \frac{-\tan\theta_W}{\sqrt{3}} \left( W_{8\mu} - \sqrt{2} W_{15\mu} \right) + \sqrt{1 - \tan^2\theta_W}\, B_\mu \right]$$

$$Z'_\mu = \sqrt{1 - \frac{\tan^2\theta_W}{3}} W_{8\mu} + \frac{\tan\theta_W}{\sqrt{3}} \frac{\sqrt{2(1-\tan^2\theta_W)}}{\sqrt{3 - \tan^2\theta_W}} \left( W_{15\mu} \frac{\tan\theta_W}{\sqrt{1-\tan^2\theta_W}} + \sqrt{\frac{3}{2}} B_\mu \right) \tag{58}$$

The eigenstates correspond to the 3-4-1 model and are same for $A_\mu, Z_\mu$ in four Higgs 4-plet model [17]. The new gauge boson $Z'_\mu$ is different in the present case.

In 3-3-1 model [18] with two-Higgs triplets, $W_{4\mu}$ mixes with $W_{3\mu}, W_{8\mu}$ only for finite u value.

We can consider mixing of $Z'$ with new gauge boson $W_4$ for non-zero VEV u value. [18,20]. This problem and its phenomenological implications is beyond the scope of this work and will be considered separately.

For $u \ll V$ no mixing occurs with $W_4$. The $K^{0*}$ mass is obtained as

$$M^2_{K^0} = \frac{g^2}{4}\left(u^2 + V^2\right) = M^2_{K_1} - M^2_W \tag{59}$$

This result is also obtained in 3-3-1 model with two-Higgs triplets [18] and three Higgs triplet model [21].

## VI. Charged currents and neutral current

### VIA. Charged currents

The Hamiltonian for the charged currents for $SU(4)_L$ is

$$H_{CC} = \frac{g}{\sqrt{2}} \left( W^+_\mu J^{-\mu}_w + Y^+_\mu J^{-\mu}_Y + K^-_{1\mu} J^{+\mu}_{K_1} + X^+_\mu J^{-\mu}_X + K^0_\mu J^{0*\mu}_K + V^0_\mu J^{0*\mu}_V + h.c. \right) \tag{60}$$

The charged current has additional terms due to mixing of $W'^\pm_\mu, K'^\pm_{1\mu}$ with mixing angle given by $\tan\beta = \frac{u}{V}$ in eqn.(50).

$$J_W^{-\mu} = \sum_{ij}[\cos\beta(\bar{u}_{iL}\gamma^\mu d_{iL} + \bar{v}_{jl}\gamma^\mu e_{jL}) + \sin\beta(\bar{U}_L\gamma^\mu d_{iL} + \bar{u}_{iL}\gamma^\mu D_{iL} + \bar{N}_{jL}\gamma^\mu e_{jL})]$$
$$+\cos\beta(\bar{t}_L\gamma^\mu b_L) + \sin\beta(\bar{t}_L\gamma^\mu D_{iL}) \tag{61}$$

$$J_{K_1}^{-\mu} = \sum_{ij}[-\sin\beta(\bar{u}_{iL}\gamma^\mu d_{iL} + \bar{v}_{jl}\gamma^\mu e_{jL}) + \cos\beta(\bar{U}_L\gamma^\mu d_{iL} + \bar{u}_{iL}\gamma^\mu D_{iL} + \bar{N}_{jL}\gamma^\mu e_{jL})]$$
$$-\sin\beta(\bar{t}_L\gamma^\mu b_L) + \cos\beta(\bar{t}_L\gamma^\mu D_{iL}) \tag{62}$$

$$J_Y^{-\mu} = \sum_i \bar{U}_{iL}\gamma^\mu D_{iL} + \bar{U}_L\gamma^\mu D_L + \sum_j \bar{N}_{jL}\gamma^\mu E_{jL} \tag{63}$$

$$J_X^{-\mu} = \sum_i \bar{U}_{iL}\gamma^\mu d_{iL} + \bar{t}_L\gamma^\mu D_L + \sum_j \bar{v}_{jL}\gamma^\mu E_{jL} \tag{64}$$

$$J_K^{0*\mu} = \sum_i \bar{D}_{iL}\gamma^\mu d_{iL} + \bar{U}_L\gamma^\mu t_L + \sum_j \bar{N}_{jL}\gamma^\mu \gamma_{jL} \tag{65}$$

$$J_V^{0*\mu} = \sum_i \bar{u}_{iL}\gamma^\mu U_{iL} + \bar{D}_L\gamma^\mu b_L + \sum_j \bar{E}_{jL}\gamma^\mu l_{jL} \tag{66}$$

**VI B. Neutral currents**

The Hamiltonian for the neutral currents $J_\mu(EM), J_\mu(Z), J_\mu(Z'), J_\mu(Z'')$ is

$$H^0 = eA^\mu J_\mu(EM) + \frac{g}{2\cos\theta_W} Z^\mu J_\mu(Z) + \frac{g}{2} Z'^\mu J_\mu(Z') + gZ''^\mu J_\mu(Z'') \tag{67}$$

where $J_\mu(EM), J_\mu(Z), J_\mu(Z'), J_\mu(Z'')$ are

$$J_\mu(EM) = \sum_f Q_f \bar{f}\gamma_\mu f ; \tag{68}$$

$$J_\mu(Z) = \sum_f \bar{f}\gamma_\mu(v_Z^f + a_Z^f \gamma_5)f ; v_Z^f = (T_{3L} - 2Q_f s_W^2), a_Z^f = -T_{3L.} \tag{69}$$

$$J_\mu(Z') = \sum_f \bar{f}\gamma_\mu(v_{Z'} + a_{Z'}\gamma_5)f$$

$$= \sum_f \bar{f}\gamma_\mu \left\{ \frac{2t_W^2}{\sqrt{3-t_W^2}}\left(\frac{T_{15L}}{\sqrt{6}} + X\right) - \sqrt{3-t_W^2}\, T_{8L} + \gamma_5\left(-\frac{2t_W^2}{\sqrt{6}\sqrt{3-t_W^2}} T_{15L} + \frac{\sqrt{3-t_W^2}}{\sqrt{3}} T_{8L}\right) \right\} f$$

(70)

$$J_\mu(Z'') = \sum_f \bar{f}\gamma_\mu(v_{Z''} + a_{Z''}\gamma_5)f$$

$$= \sum_f \bar{f}\gamma_\mu \left\{ \frac{\sqrt{1-t_W^2}}{\sqrt{3-t_W^2}}\left(\sqrt{3}T_{15L} - 2\sqrt{2}X\frac{t_W^2}{1-t_W^2} - \gamma_5\sqrt{3}T_{15L}\right) \right\} f \qquad (71)$$

## VII: Results and conclusions

In this work we have presented the phenomenology of the 3-4-1 model without exotic charges for the case of a = 1, b = -1 and c = 2 parameters in the electric charge operator. The model is found to be embedded in $SU(3)_C \otimes SU(4)_L \otimes U(1)_{B-L} \otimes U(1)_{Y_0}$ which is a subgroup of SU(8) gauge group. The scalar sector consists of two-Higgs 4-plet and a singlet of $SU(4)_L$ scalars and is studied in detail to give nine physical Goldstone bosons, three massive neutral scalars and one charged massive Higgs scalar. There are no massive pseudoscalars so that there is no CP violation in the Higgs sector. The fermion spectrum has additional massive up and down types of quarks and charged leptons. The gauge boson sector contains two additional new neutral currents .There are two charged gauge bosons, one of which can mix with the SM gauge field. There is possible mixing between neutral gauge bosons .This mixing is model dependent and determines the mass of new neutral gauge boson $Z'$.

The additional $U(1)_{Y_0}$ symmetry is a special feature of this model and determines the X charge which is fixed by anomaly cancellations in all other 3-4-1 models. The $Y_0$ charge is $\pm\frac{1}{2}$ for Higgses and SU(4)$_L$ singlet fermions and zero for all other states. This is different from the U(1) gauge symmetries in E$_6$ derived GUT models [22] as well as string-derived $U(1)_X$ symmetries .[23] which predict additional neutral gauge boson $Z'$. In the present 3-4-1 model, X charge is given by eqn.2 as $\left(\frac{B-L}{2}+Y_0\right)$ so that $B_\mu$ (eqn.42) corresponds to $U(1)_X$ gauge boson.

Additional scalar singlet $\phi_0$, introduced for symmetry breaking, has lepton number L = -2.

The VEV's satisfy the condition u << $v, V, z$. The model has interesting phenomenological implications which are similar but not exactly identical to ref [18,20] 3-3-1 case as analyzed in section **IIB.** Finally we mention that 3-4-1 model without exotic charges of type F has not been studied in literature [17] and thus the present work is an interesting step in this direction.

**Acknowledgements**

Work partially supported by University Grants Commission

(Ref.No.F5.1.3(168)/2004(MRP/NRCB)